\documentstyle[11pt,psfig]{article}
\def\squareforqed{\hbox{\rlap{$\sqcap$}$\sqcup$}}
\def\sq{\ifmmode\squareforqed\else{\unskip\nobreak\hfil
\penalty50\hskip1em\null\nobreak\hfil\squareforqed
\parfillskip=0pt\finalhyphendemerits=0\endgraf}\fi}
\def\arcmin{\hbox{$^\prime$}}
\def\arcsec{\hbox{$^{\prime\prime}$}}

\def\farcs{\hbox{$.\!\!^{\prime\prime}$}}
\def\la{\mathrel{\mathchoice {\vcenter{\offinterlineskip\halign{\hfil
$\displaystyle##$\hfil\cr<\cr\sim\cr}}}
{\vcenter{\offinterlineskip\halign{\hfil$\textstyle##$\hfil\cr
<\cr\sim\cr}}}
{\vcenter{\offinterlineskip\halign{\hfil$\scriptstyle##$\hfil\cr
<\cr\sim\cr}}}
{\vcenter{\offinterlineskip\halign{\hfil$\scriptscriptstyle##$\hfil\cr
<\cr\sim\cr}}}}}
\def\ga{\mathrel{\mathchoice {\vcenter{\offinterlineskip\halign{\hfil
$\displaystyle##$\hfil\cr>\cr\sim\cr}}}
{\vcenter{\offinterlineskip\halign{\hfil$\textstyle##$\hfil\cr
>\cr\sim\cr}}}
{\vcenter{\offinterlineskip\halign{\hfil$\scriptstyle##$\hfil\cr
>\cr\sim\cr}}}
{\vcenter{\offinterlineskip\halign{\hfil$\scriptscriptstyle##$\hfil\cr
>\cr\sim\cr}}}}}
\newcommand{\sbb}{mag/$\sq{\rm ''}$}
\def\ko{K\"O2000}
\def\ha{H$\alpha$}

\title{{\bf Optical \& NIR surface photometry of I Zw 18}\footnote{
Research by P.P, K.G.N and K.J.F. has been supported by DARA GmbH 
grant 50\ OR\ 9907\ 7 and DFG grant FR 325/50--1.
Y.I.I. thanks for a Gau\ss\ professorship of the G\"ottingen Academy of
Sciences. We acknowledge the financial support of the Volkswagen Foundation Grant 
No. I/72919. We thank the Calar Alto staff for their assistance during the observations.}}
\author{P. Papaderos$^1$, Yu.I. Izotov$^2$, K.G. Noeske$^1$, T.X. Thuan$^3$, K.J. Fricke$^1$
\vspace{0.1cm}\\
\normalsize
$^1$ Universit\"ats--Sternwarte, Geismarlandstra\ss e 11,
                 D--37083 G\"ottingen, Germany\\
$^2$ Main Astronomical Observatory,
                 Ukrainian National Academy of Sciences,
                 Kyiv 03680,  Ukraine\\
$^3$ Astronomy Department, University of Virginia, 
                 Charlottesville, VA 22903, USA}
\date{}
\begin{document}
\maketitle
\def\bull{\vrule height .9ex width .8ex depth -.1ex}
\makeatletter
\def\ps@plain{\let\@mkboth\gobbletwo
\def\@oddhead{}\def\@oddfoot{\hfil\tiny
``Dwarf Galaxies and their Environment'';
Bad Honnef, Germany, 23-27 January 2001; Eds.{} K.S. de Boer, R.-J.Dettmar, U. Klein; Shaker Verlag}%
\def\@evenhead{}\let\@evenfoot\@oddfoot}
\makeatother

\begin{abstract}\noindent
Using {\sl HST} and ground-based optical and NIR data,
we investigate whether the blue compact dwarf (BCD) galaxy I Zw 18 has 
an extended low-surface-brightness (LSB) older stellar population underlying  
the star-forming regions, as is the case in evolved iE/nE BCDs.
Subtraction of narrow band \ha\ and [O III] exposures from 
$R$ and $V$ images shows that the filamentary LSB envelope 
extending out to $\sim$2 kpc away from the starburst region, and hence
the optical broad-band colors observed therein, are due mainly 
to ionized gas emission.
Ionized gas accounts already at a galactocentric distance of
0.7 kpc for more than 80\% of the $R$ band line-of-sight intensity
and contributes more than 40\% of the integrated $R$ band light of I Zw 18. 
The structural properties (such as the exponential scale length) of the
{\it stellar} LSB component underlying the extended ionized gas emission
place I Zw 18 among the most compact BCDs studied so far.
Contrary to evolved nE/iE BCDs the stellar 
component in I Zw 18 shows no appreciable color gradients over 
a range of $\sim$8 mag in surface brightness.
\end{abstract}
In order to understand the dynamical formation and the 
evolutionary state of I Zw 18, it is important to investigate whether or not
the star-formation activity in this system is taking place 
on top of an extended and evolved {\it stellar} low-surface-brightness (LSB)
host. 
If the putative underlying older stellar population has  
photometric properties typical of evolved (a few Gyr old) 
iE/nE BCDs (cf. Loose \& Thuan 1985), then 
its detection should be feasible with current instrumentation.
The underlying LSB component of BCDs shows for $M_B\ga$--16 mag  
a central surface brightness excess of $\ga$1.5 mag and 
an exponential scale length reduced by a factor 
of $\sim$2 as compared to, e.g., 
dwarf irregulars (Papaderos et al. 1996; hereafter P96, Patterson \& Thuan 
1996, Salzer \& Norton 1999), i.e. its mean surface
brightness is significantly higher than in other dwarf galaxy types.
Moreover, in such systems the emission of the evolved stellar host contributes 
on average about
half of the $B$ band light within the 25 $B$ \sbb\ isophote and dominates the
radial intensity and color distribution for fainter surface brightness levels (P96).

In the following, we investigate the photometric properties and nature (stellar
and/or gaseous) of the extended LSB envelope surrounding I Zw 18 (cf. Fig. 1a\&c) using ground-based and {\sl HST} data. We 
adopt a distance of 15 Mpc 
(Izotov et al. 2001) and a mean extinction $A_V$=0.16 mag for I Zw 18.
This study is based on archival broad-band {\sl HST} WFPC2 data in 
$B$ (F450W), $V$ (F555W) and $R$ (F702W) by Dufour et al. (1996). 
Narrow band [O III] (F502N) and H$\alpha$ (F658N) images were used 
to correct the broad band data for the contribution of ionized gas line emission.
Furthermore, we include 
$B$ (90 min), $V$ (40 min), $R$ (70 min) and $I$ (60 min) ground-based data
taken with the Calar Alto 1.23m/2.2m and Kitt Peak 2.1m telescopes.
NIR exposures in $J$ (66 min), $H$ (36 min) and $K$\arcmin\ (35 min) were 
acquired with the Omega 1k$\times$1k camera attached to the prime focus 
of the 3.5m telescope at Calar Alto. 
\begin{figure}[t]
\begin{picture}(16.4,6.3)
\put(5.2,3){please download 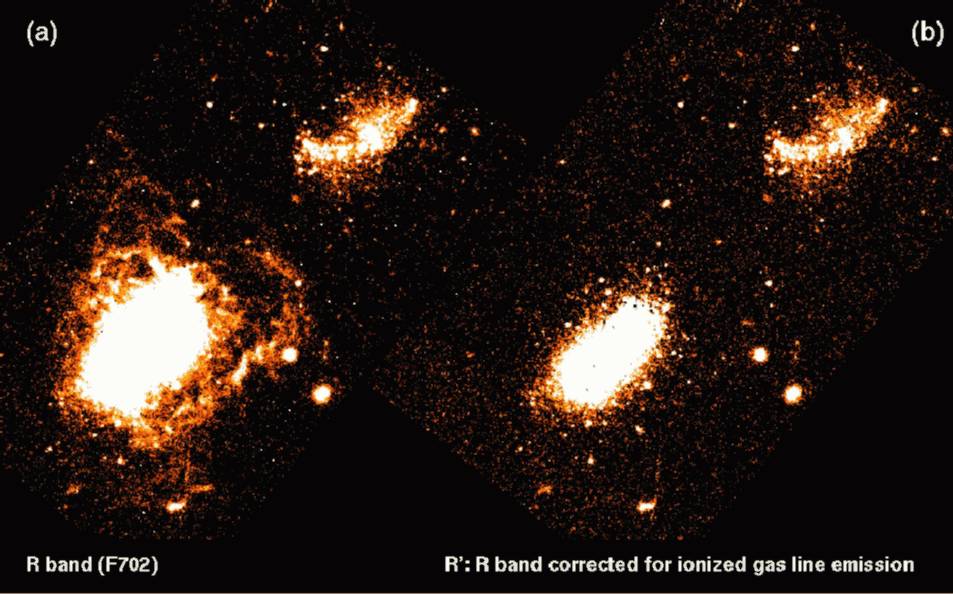 and 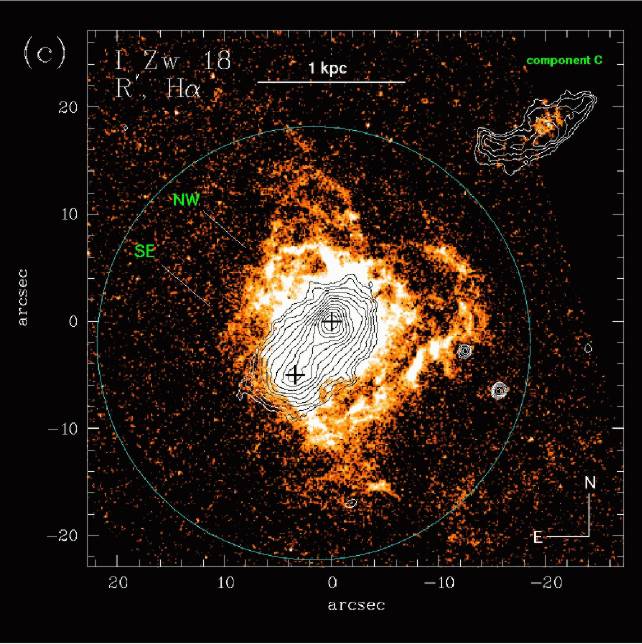}
\put(4.8,2.5){or retrieve the original article from the URL :}
\put(3.5,2){http://www.astro.uni-bonn.de/$\sim$webgk/dgeproc/dge111.ps.gz }
\end{picture}
\caption[]{{\bf (a)} {\sl HST} WFPC2 exposure of I\ Zw\ 18 in the $R$ (F702W) 
band. {\bf (b)} The same $R$ band exposure after scaling and subtraction of 
the \ha\ (F658N) map (in the following referred to as $R$\arcmin). 
Both images are displayed in the same intensity range.
{\bf (c)} $R$\arcmin\ contours overlaid with an \ha\ map of the 
main body of I Zw 18. Contours correspond to surface brightness levels
from 18.5 $R$\arcmin\ \sbb\ to 25 $R$\arcmin\ \sbb\ in steps of 0.5 mag.
The SE and NW star-forming regions are indicated by crosses.} 
\end{figure}

The surface brightness profiles (SBPs) of the main body 
of I\ Zw\ 18, derived from ground-based $B$ and $R$ data (Fig. 2a), 
can be well fitted by an exponential law 
for radii 8\arcsec$\leq$ $R$$^*\la$16\arcsec. 
A linear fit to the $B$ band profile yields for the LSB envelope 
a central surface brightness $\mu_0=22.3\pm0.1$ $B$ \sbb, an exponential 
scale length of 3\farcs 45 ($\sim$250 pc) and an apparent magnitude of 
$m_B$=17.7 mag ($\sim$24\% of the total $B$ band emission). 
\begin{figure}[!h]
\begin{picture}(16.4,10)
\put(0,5.25){{\psfig{figure=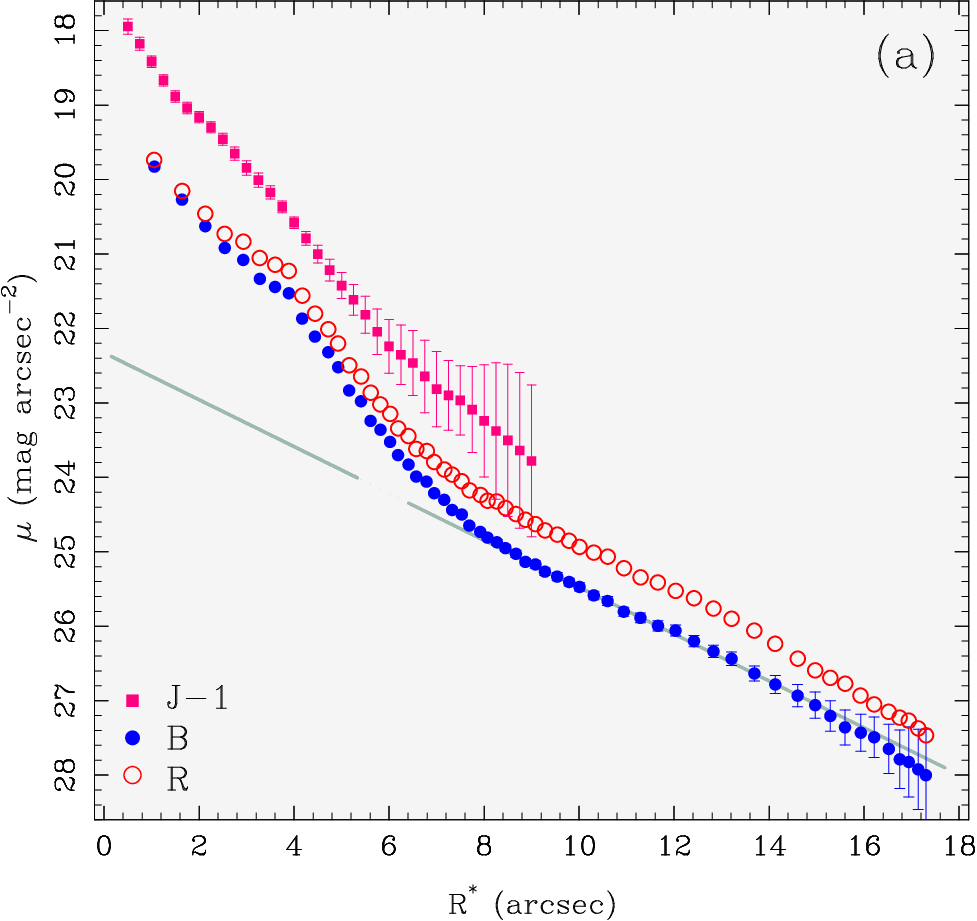,height=5.cm,angle=-90.,clip=}}}
\put(5.7,5.25){{\psfig{figure=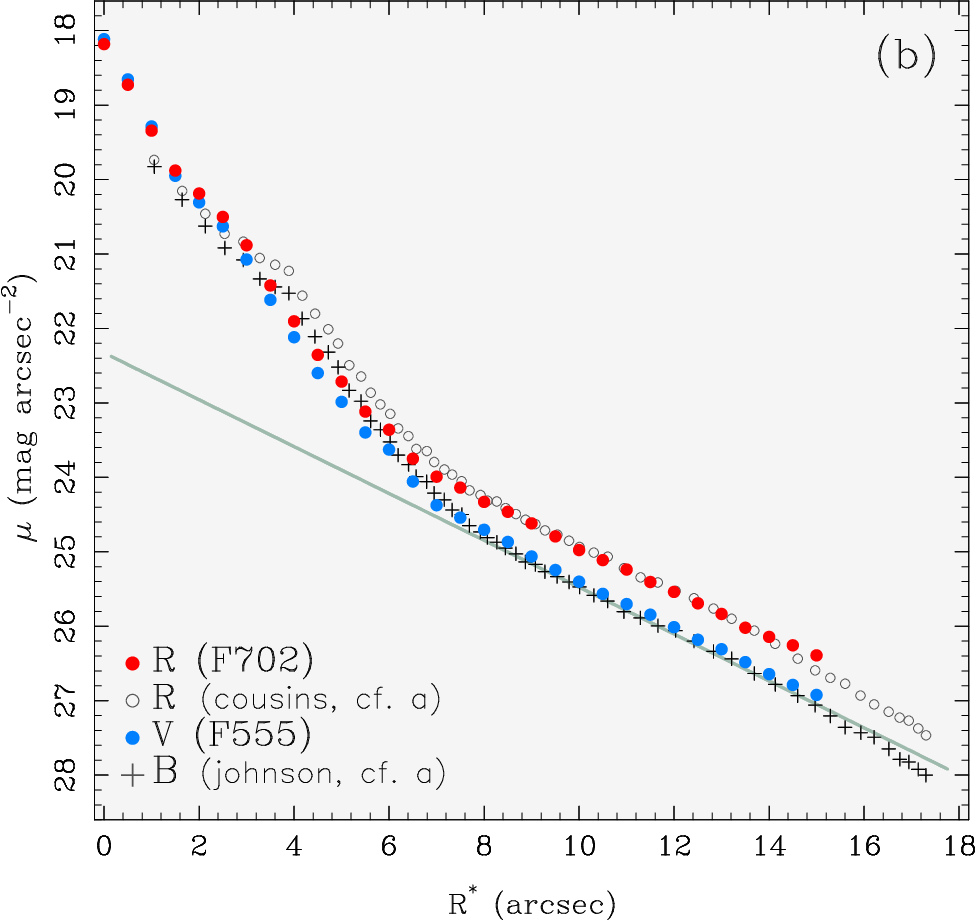,height=5.cm,angle=-90.,clip=}}}
\put(11.4,5.25){{\psfig{figure=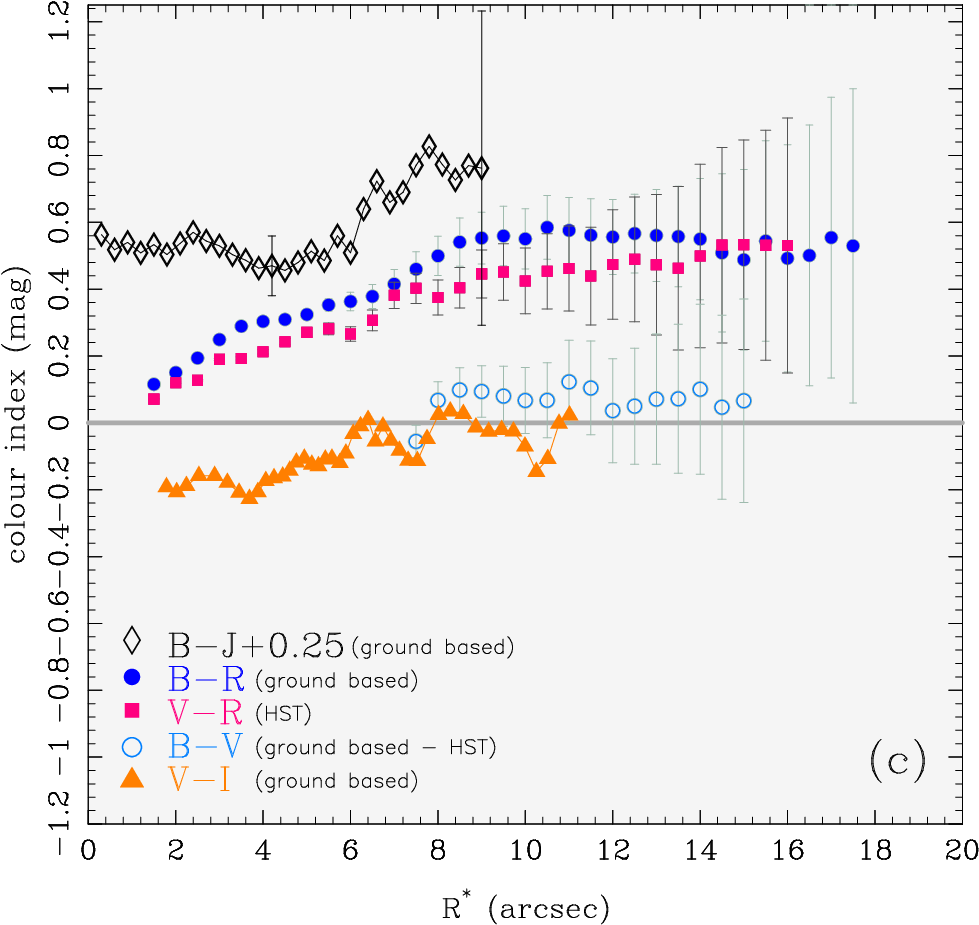,height=5.cm,angle=-90.,clip=}}}
\put(0,0){{\psfig{figure=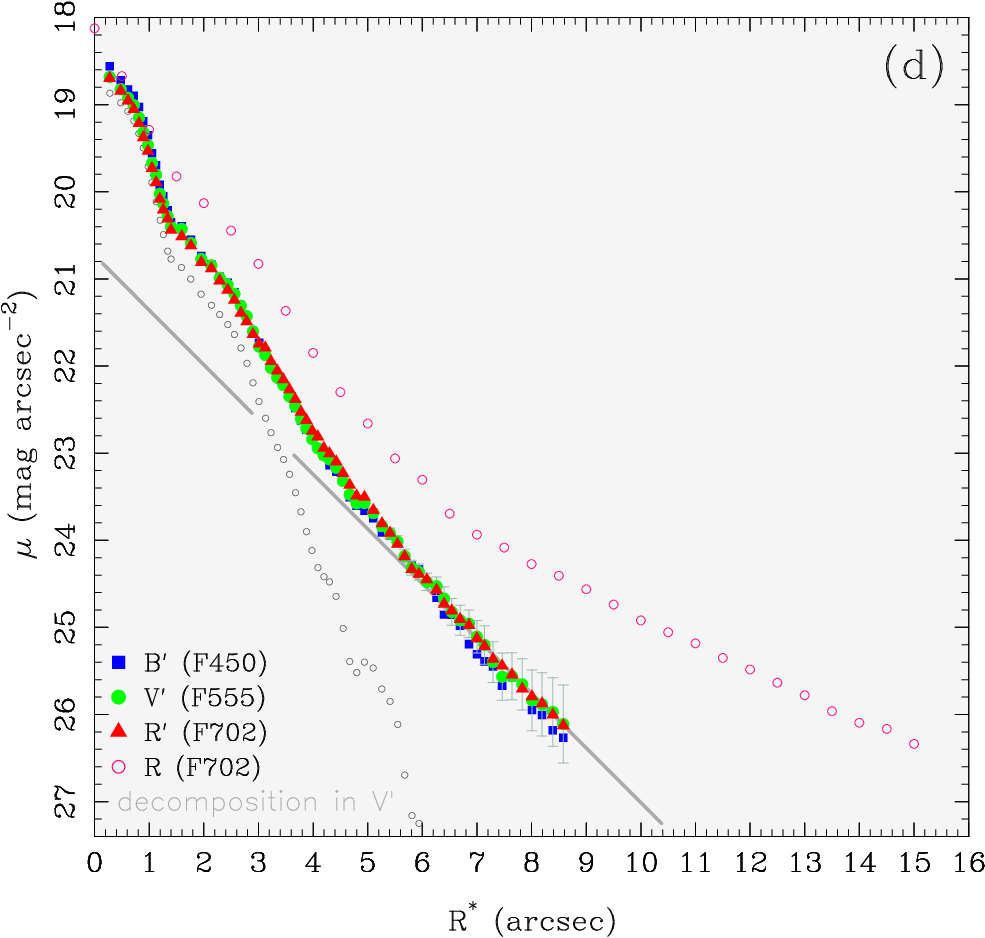,height=5.cm,angle=-90.,clip=}}}
\put(5.7,0){{\psfig{figure=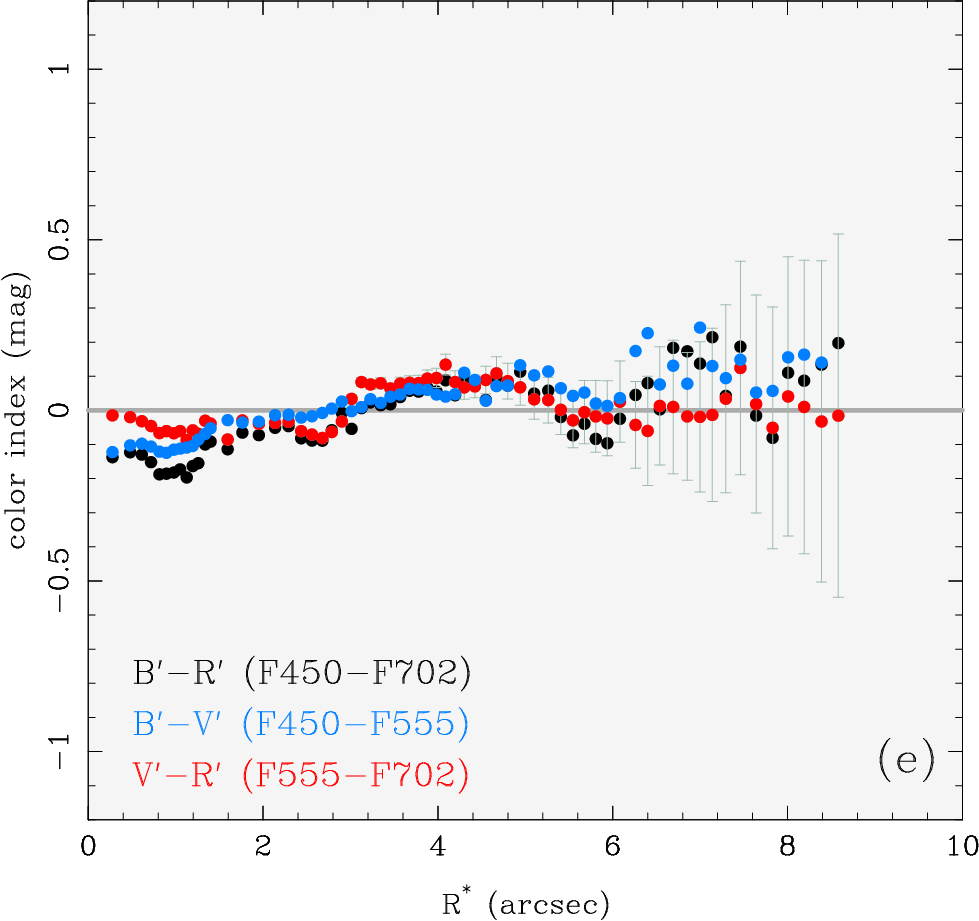,height=5.cm,angle=-90.,clip=}}}
\put(11.4,0){{\psfig{figure=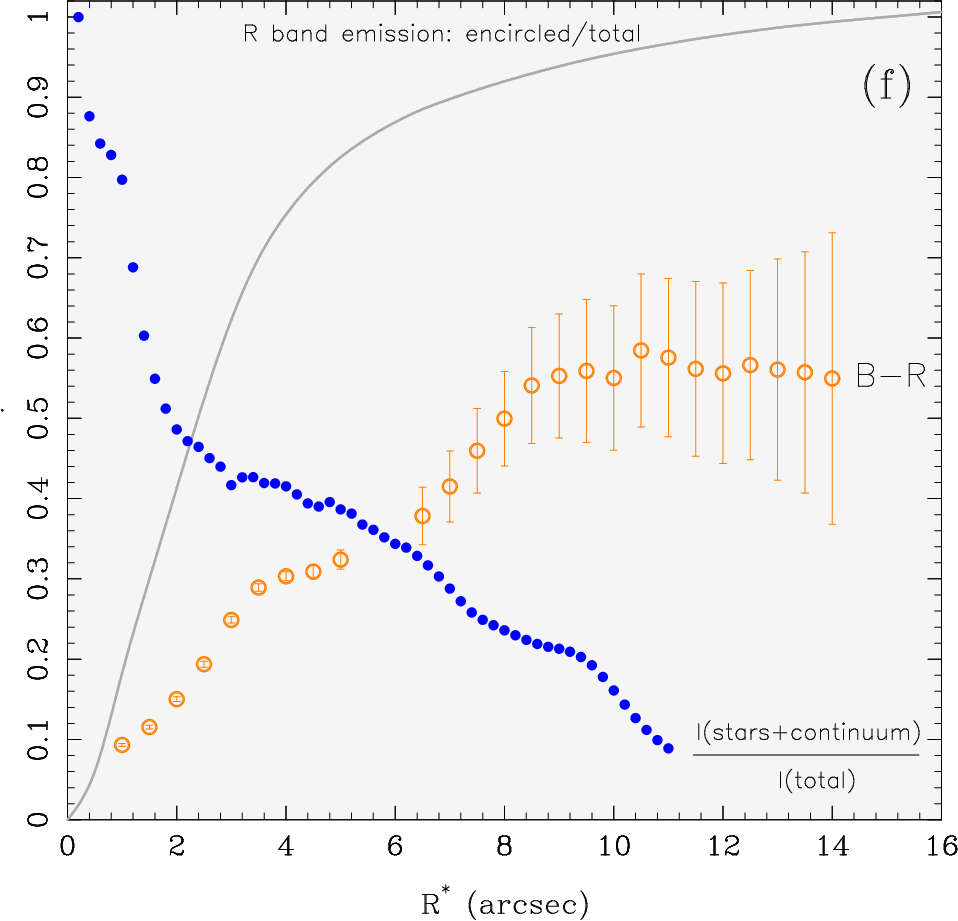,height=5.cm,angle=-90.,clip=}}}
\end{picture}
\caption[]{{\bf (a)} Surface brightness profiles (SBPs) of I Zw 18's main body
in $B$, $R$ and $J$ computed from ground-based data. The straight line shows a 
fit to the exponential regime of the $B$ band SBP for radii $\geq$9\arcsec.
{\bf (b)} Comparison of the $B$ and $R$ SBPs shown in panel (a) with those
obtained from archival {\sl HST} WFPC2 data in $V$ (F555W) and $R$ (F702W). 
Note that for radii $>$9\arcsec\ the $B$ band SBP matches closely 
the $V$ band profile.
{\bf (c)} $B-R$, $V-R$, $V-I$, $B-V$ and $B-J$ color indices
computed from ground-based and {\sl HST} data.
{\bf (d)} SBPs of I\ Zw\ 18 in $B$, $V$ and $R$ computed
after correction for the contribution of ionized gas line emission 
(labelled $B$\arcmin, $V$\arcmin, $R$\arcmin). 
A linear fit to the $V$\arcmin\ band SBP for $R$$^*\geq$6\farcs5 is 
shown by the solid line. The SBP in the $R$ band as computed before 
correction for the \ha\ emission (cf. panel a) is included for comparison.
{\bf (e)}  Color profiles of the main body of I Zw 18
derived from SBPs in $B$\arcmin, $V$\arcmin\ and $R$\arcmin.
{\bf (f)}
Fractional contribution of the stellar background and the gaseous continuum 
to the $R$ band line-of-sight intensity of the main body of I Zw 18 as 
a function of the equivalent radius $R^*$ (filled circles). 
For $R^*>$9\farcs6 ionized gas accounts for 
$>$80\% of the $R$ band emission and hence determines the moderately 
red $B-R$ color (open symbols) observed all over the filamentary 
LSB envelope of the BCD.}
\end{figure}

The same result can be verified independently from {\sl HST} data. 
As shown in Fig. 2b, {\sl HST} WFPC2 data in $V$ (F555W) and $R$ (F702W) 
allow to trace the surface brightness distribution of the LSB host 
out to its Holmberg radius. The SBPs derived from the latter data
are in the radial range 8\arcsec$\leq R^*\leq$14\arcsec\ practically 
indistinguishable from those inferred from ground-based images in $B$ and $R$.
This lends further support to the conclusion of Papaderos (1998) 
and Kunth \& \"Ostlin (2000; hereafter \ko) that the surface 
brightness distribution of the LSB envelope of I Zw 18,
despite its patchiness, can be well approximated by an exponential fitting law.

The $B-R$ and $V-R$ colors of I Zw 18 (Fig. 2c) increase smoothly with
galactocentric distance with a mean gradient of $\sim$0.05 mag/\arcsec\ 
and level off at $B-R\approx$+0.55 mag and $V-R\approx$+0.47 mag for $R^*>$8\arcsec. 
This $B-R$ color derived for the outlying regions of I Zw 18 is
in good agreement with the value of $\sim$+0.6 mag 
obtained by Papaderos (1998) and \ko. 
The $B-J$ color index was determined to be $\la$+0.6 mag at the
equivalent radius of $R^*=9$\arcsec. 
Beyond that radius, corresponding to a surface brightness level 
of $\sim$25 \sbb\ in the $J$ band, noise and source confusion do not allow
a reliable determination of optical--NIR colors.
We note that our data {\sl do not show} the steep and systematic 
increase of the $B-J$ index (up to +1.7 mag at $R^*$$\approx$12\arcsec) 
reported by \ko.

Assuming that ionized gas emission is negligible in the outskirts 
of I Zw 18, then the relatively red $B-R$ or $V-R$ colors 
would imply that the extended LSB component of 
this system is made of an evolved stellar population, typical for iE/nE BCDs.
Indeed, evolutionary synthesis models for the colors of this component
would yield an age as high as $\sim$20 Gyr.
However, the assumption that the LSB emission in I Zw 18 is primarily 
 of stellar origin is not tenable when all  
the color distributions displayed in Fig. 2c are considered.
The $B-V$ and $V-I$ indices remain blue ($\sim$0 mag) out 
to $R^*\approx$14\arcsec, i.e. over a range of 8 mag 
in surface brightness.
No model with only a stellar population and no gaseous emission 
can reproduce the colors observed in the outskirts
of I Zw 18 (e.g. red $V-R$ and blue $V-I$ colors), 
irrespective of the age or metallicity of that stellar population. 

The striking filamentary appearance of the LSB envelope of I Zw 18 (Fig. 1a,c),
together with its observed colors both suggest a substantial contribution 
of ionized gas emission for radii $\ga$4\arcsec\ (cf. eg. Izotov et al. 1997, 
Papaderos et al. 1998).
\"Ostlin et al. (1996) first presented an \ha\ equivalent width ($EW$(\ha))
map of I Zw 18 from high resolution ground-based data and inferred 
an $EW$(\ha) as high as $\sim$ 1500 \AA\ over an appreciable fraction of its
main body. The significant contribution of ionized gas emission in the 
halo of I Zw 18 has also been discussed by 
V\'{\i}lchez \& Iglesias-P\'{a}ramo (1998) and Izotov et al. (2001).

An upper limit to the isophotal size and scale length
of the stellar component in I Zw 18 can be obtained by
scaling and subtracting [O\,III] and \ha\ maps from the 
$B$(F450W)\footnote{Note that the transmission curve of
the F450W filter includes the [O III]$\lambda$5007 emission line.}, 
$V$(F555W) and $R$(F702W) images. 
The resulting images, referred to as 
$B$\arcmin, $V$\arcmin\ and $R$\arcmin, account therefore mainly
for the combined emission of the stellar background 
and the gaseous continuum in the main body of the BCD.
Figure 1b illustrates the importance of this correction 
for ionized gas line emission to the study of the stellar
content in I Zw 18: while subtraction of ionized gas emission
has practically no effect on the morphology and photometric properties
of component C where the \ha\ line is very weak,
it results in the virtual removal of the filamentary
LSB envelope surrounding the main body.

The photometric properties of the compact (the effective radius is 
$\sim$150 pc) and relatively smooth residual emission can be derived from the $B$\arcmin, 
$V$\arcmin\ and $R$\arcmin\  surface brightness profiles (Fig. 2d).
Linear fits for $R$$^*\geq$6\farcs5 yield consistently a mean 
exponential scale length of 125$\pm$10 pc and a $V$\arcmin\ band 
central surface brightness of 20.7$\pm$0.4 \sbb, 
placing I Zw 18 among the most compact BCDs studied so far. 
The narrow range of scale lengths in all these bands 
implies a nearly constant color throughout the main 
body of I Zw 18. 
The color profiles derived from $B$\arcmin, $V$\arcmin\ 
and $R$\arcmin\ data (Fig. 2e) are still contaminated  
by line and continuum emission from ionized gas,  
so their interpretation in terms of stellar population age 
is not straightforward.
It is worth noting, however, that color profiles computed this way
show -- contrary to those displayed in panel c -- a weak gradient and 
a small scatter of $\sim$0.2 mag around zero.

From profile integration, we obtain a lower limit for
the contribution of gaseous emission to the integrated 
light of I Zw 18 of $\sim$20\% and $\sim$40\% in $V$ and $R$, 
respectively.
The contribution of the \ha\ emission line 
to the observed intensity as a function of the equivalent radius $R$$^*$
can be assessed from the $R$ and $R$\arcmin\ SBPs (Fig. 2a and Fig. 2d).
Figure 2f shows that for $R$$^*>$9\farcs6 (0.7 kpc) the average
contribution of the stellar background to the $R$ band line-of-sight intensity
is less than 20\%, implying that the colors observed in the LSB envelope of 
I Zw 18 (cf. Fig. 2c) result mainly from ionized gas emission.

For larger radii, the available data do not go deep enough to 
allow to put firm constraints on the surface density of 
the stellar background possibly underlying the ionized gas 
envelope of I Zw 18.
An estimate can be obtained by extrapolating the exponential
slope determined for the $B$\arcmin, $V$\arcmin, $R$\arcmin\ SBPs (Fig. 2d). 
Within the 2$\sigma$ bound in the profile fitting,  
we obtain upper limits of 21.6 $B$ \sbb\ for the central surface brightness,
and of 146 pc for the scale length of the underlying stellar component in 
I Zw 18. These values yield for the surface brightness of the stellar 
component a conservative upper limit of $\sim$30.5 $B$\arcmin\ \sbb\ 
at a galactocentric distance of 1.2 kpc (16\farcs5), i.e. 
only $\sim$7\% of the observed $B$ band intensity.

{\small
\begin{description}{} \itemsep=0pt \parsep=0pt \parskip=0pt \labelsep=0pt
\item {\bf References}
%
\item Dufour, R.J., Garnett, D.R., Skillman, E.D., Shields, G.A. 1996, {\sl From Stars To Galaxies}, ASP Conference
  Series 98, eds. C. Leitherer, U. Fritze-v. Alvensleben, J. Huchra, p.358
\item Izotov, Y.I., Lipovetsky, V.A., Chaffee, F.H., Foltz, C.B., Guseva, N.G.,
  Kniazev, A.Y. 1997, ApJ 476, 698
\item Izotov, Y.I. et al. 2001, in prep.
\item Loose, H.-H., Thuan, T.X. 1985, in {\sl Star-Forming Dwarf Galaxies},
  eds. D. Kunth, T.X. Thuan, T.T. Van, Editions Fronti\`eres, p. 73
\item Kunth, D., \"Ostlin, G. 2000, A\&AR 10, 1 (\ko)
\item \"Ostlin, G., Bergvall, N., R\"onnback, J. 1996, {\sl The Interplay 
Between Massive Star Formation, the ISM and Galaxy Formation}, eds. D. Kunth, et
al., Editions Fronti\`eres, p. 605
\item Papaderos, P., Loose, H.-H., Fricke, K.J., Thuan, T.X. 1996, A\&A 314, 59
\item Papaderos, P. 1998, PhD thesis, Universit\"at G\"ottingen
\item Papaderos, P., Izotov, Y.I., Fricke, K.J., Thuan, T.X., Guseva, N.G. 1998,
  A\&A 338, 43
\item Patterson, R.J., Thuan, T.X. 1996, ApJS 107, 103
\item Salzer, J.J., Norton, S.A. 1999, IAU Colloquium 171
\item V\'{\i}lchez, J. M., Iglesias-P\'aramo J. 1998, ApJ 508, 248
\end{description}
}
\end{document}